\documentclass[12pt,a4paper]{article}
\usepackage[latin1]{inputenc}
\usepackage{graphicx}
\usepackage{multirow}
\usepackage{boldline}
\usepackage{subfig}
\usepackage[left=20.00mm, right=20.00mm, top=20.00mm, bottom=20.00mm]{geometry}
\author{Muhammad Abdul Wahab$~^\alpha$, Pascal Cotret$~^\beta$, Mounir Nasr Allah$~^\delta$\\Guillaume Hiet$~^\delta$, Arnab Kumar Biswas$~^\gamma$, Vianney Lapôtre$~^\gamma$, Guy Gogniat$~^\gamma$\\~\\
$~^\alpha$ IETR/SCEE research group, firstname.lastname@centralesupelec.fr\\
$~^\beta$ Independent researcher, pascal.cotret@gmail.com\\
$~^\delta$ INRIA/CIDRE research group, firstname.lastname@centralesupelec.fr\\
$~^\gamma$ Lab-STICC/University of South Brittany, firstname.lastname@univ-ubs.fr}
\title{\vspace{-4em}A small and adaptive coprocessor for information flow tracking in ARM SoCs}
\date{(Preprint version)}
\begin{document}
\maketitle
\begin{abstract}
DIFT (\emph{Dynamic Information Flow Tracking}) has been a hot topic for more than a decade. Unfortunately, existing hardware DIFT approaches have not been widely used neither by research community nor by hardware vendors. It is due to two major reasons: current hardware DIFT solutions lack support for multi-threaded applications and implementations for hardcore processors. This work addresses both issues by introducing an approach with some unique features: DIFT for multi-threaded software, virtual memory protection (rather than physical memory as in related works) and Linux kernel support using an information flow monitor called RFBlare. These goals are accomplished by taking advantage of a notable feature of ARM CoreSight components (context ID) combined with a custom DIFT coprocessor and RFBlare. The communication time overhead, major source of slowdown in total DIFT time overhead, is divided by a factor 3.8 compared to existing solutions with similar software constraints as in this work. The area overhead of this work is lower than 1\% and power overhead is 16.2\% on a middle-class Xilinx Zynq SoC.
\end{abstract}

\section{Introduction}\label{introduction}
IFT (\emph{Information Flow Tracking}) consists of adding tags to information containers, propagating and checking these tags during program execution. A tag is a value associated with the information container and represents a level of security. For instance, if only two tag values are considered, the tags can represent private or public information (a tag with $N$ bits can represent $2^N$ security levels). When containers are processed in a program, tags must be updated: this step is known as \emph{tags computation}. IFT can be used to detect or prevent software attacks such as buffer overflows, SQL injection or confidential data leakage \cite{Dalton2007}.
There are two types of IFT:
\begin{itemize}
	\item SIFT  (\emph{Static IFT}) consists of an off-line analysis of the binary to check that all branches of the program are trustworthy.
	\item DIFT (\emph{Dynamic IFT} is performed at runtime: it monitors data flow of the program to make sure that no unauthorized operation is done in the current execution branch. 
\end{itemize}

SIFT is an appealing solution since it does not introduce any runtime overhead and ensure that the program is safe before executing it. However, this approach is not suited to protect a whole system consisting of different complex applications developed using different languages. DIFT is a more practical and flexible solution that can take into account some information that is hard to predict before execution, such as dynamic memory allocation, data or code that depends on the program inputs, etc. In this article, we propose a hybrid approach. The DIFT monitor described in this work relies on annotations that were statically pre-computed at compile time.

First, DIFT has been implemented in software but the performance overhead of such solutions is their main drawback (the application runs 37 times slower than without DIFT \cite{Dalton2007}). In the last decade, multiple architectures taking advantage of FPGAs have been proposed to reduce this performance overhead. This improvement comes at the expense of flexibility provided by software IFT. Moreover, there is currently no solution provided by CPU hardware vendors (for instance, ARM and Intel) that allows implementing DIFT. Solutions like ARM TrustZone or Intel SGX allow protecting memory regions and provide security features for trusted applications. However, no security guarantees are provided if applications are untrusted, which is the case in this work.\\

On the one hand, most of related works implement DIFT on softcores \cite{Dalton2007,Kannan2009,Heo2015,Dhawan2015} allowing to test the approach quickly. But most of these architectures are not directly portable to hardcore CPUs. DIFT requires information at runtime, such as the current executed instruction or accessed memory addresses, which can be easily recovered from softcore CPUs through existing signals.

On the other hand, other works such as\cite{Wahab2017} rely on ARM-based SoCs which do not export enough information from the CPU. The common way of recovering information for DIFT on a hardcore CPU is through instrumentation. Instructions are added in the program in order to recover information that is used to compute and check tags. However, the performance overhead due to instrumentation is high: it can represent up to 90\% of the total DIFT overhead~\cite{Heo2015}. Furthermore, the solution proposed in \cite{Heo2015} is still based on a softcore CPU which is not the final target of this work.\\ 

Debug components can also be used to retrieve such information~\cite{Wahab2017}. Depending on the type of debug components, the program may still need instrumentation to recover missing information. ARM CPUs include two debug components: ETM and PTM. If an ETM (\emph{Embedded Trace Macrocell}) \cite{ARMCoreSightETM} with data trace component is included on the device, there is no need to instrument the code. The ETM sends an information for each CPU instruction executed. However, CPU that target performance-intensive systems using rich OS (\emph{Operating System}), such as Cortex-A cores, use the PTM (\emph{Program Trace Macrocell} \cite{Wahab2017}). This debug component, considered in this work, only sends information about branch or jump addresses. This solution requires to statically analyze and instrument the program to recover missing information in the PTM trace \cite{Wahab2017}.\\

This work targets ARM hardcore CPUs running a Linux kernel. An MMU (\emph{Memory Management Unit}) takes care of translating virtual addresses to physical addresses when executing applications. Existing works associate a tag to the physical address. However, Linux applications contain virtual addresses. To resolve the difference in these address spaces, related work \cite{Lee2016} uses a lookup table to keep track of translations done by the kernel. In this work, we propose a novel solution to tag virtual memory instead of physical memory. It provides the advantage of avoiding recovery overhead of virtual to physical address translations. Existing hardware-assisted DIFT solutions lack Linux kernel support. Most solutions do not target modern operating systems. Works considering Linux OS lack the information on how to initialize the tag of information containers and other kernel modifications. This is very important because if the initialization is not done correctly, DIFT will fail in detecting attacks. Therefore, all the kernel modifications we propose are explained in this article, especially the communication between kernel and DIFT coprocessor.\\

Existing works target single-core CPUs. However, most of current CPUs are multi-core, even in embedded systems. As a consequence, the approach developed in this work is compatible with multi-threaded applications. To the best of our knowledge, no existing hardware approach implements DIFT for multi-threaded applications. This paper is organized as follows. Section \ref{related-work-and-assumptions} provides insights on existing hardware DIFT solutions. Section \ref{proposed-architecture} presents the proposed architecture and provides implementation details. Section \ref{case-studies} provides two case studies. Section \ref{implementation-results} details implementation results and Section \ref{conclusion} gives some conclusions and future perspectives.

\section{Related work and assumptions}\label{related-work-and-assumptions}
Hardware-assisted DIFT has been a hot topic for the last decade. Important works have been done to implement efficient DIFT on FPGAs. There are four main types of hardware DIFT approaches proposed in the literature~\cite{Wahab2017}:
\begin{itemize}
	\item \textbf{Filtering hardware accelerators} \cite{Fytraki2014,Chen2008}. Instead of computing tags for each executed CPU instruction, this approach proposes to eliminate unmonitored events before computing and checking tags to lower DIFT time overhead. 
	\item \textbf{In-core} approach \cite{Dalton2007,Dhawan2015} modifies the architecture of the CPU to compute tags in parallel to regular operations. 
	\item \textbf{Off-loading} approach \cite{Nagarajan2008} takes advantage of multi-core systems to offload tag computation on another general purpose core.
	\item \textbf{Off-core} approach \cite{Kannan2009,Wahab2017} consists of using a custom DIFT coprocessor to compute tags. The idea is similar to the offloading solution but instead of wasting a general purpose core, this approach uses a custom DIFT coprocessor.
\end{itemize}

Among the above solutions, only offloading and off-core DIFT approaches are portable to hardcores. The in-core solution modifies the hardware architecture and therefore is not feasible in practice. The offloading approach wastes a general purpose core for DIFT operations. Therefore, the off-core approach is the best candidate to implement DIFT for a hardcore.

\begin{table}[htbp]
	\centering
	\caption{Comparison with previous off-core approaches}
	\label{table:related_work_comparison}
	\resizebox{\textwidth}{!}{%
		\begin{tabular}{|l|l|l|l|l|l|l|l|}
			\hline
			\textbf{Approaches} & \multicolumn{1}{c|}{\textbf{Target CPU}} & \multicolumn{1}{c|}{\textbf{\begin{tabular}[c]{@{}c@{}}Multi-threaded\\ support\end{tabular}}} & \multicolumn{1}{c|}{\textbf{Tagged memory}} & \multicolumn{1}{c|}{\textbf{Kernel support}} & \multicolumn{1}{c|}{\textbf{Tag bits}} & \multicolumn{1}{c|}{\textbf{Tag scheme}} & \multicolumn{1}{c|}{\textbf{\begin{tabular}[c]{@{}c@{}}Floating-point\\ support\end{tabular}}} \\ \hline
			Kannan et al. \cite{Kannan2009} & \begin{tabular}[c]{@{}l@{}}Leon3\\ (softcore)\end{tabular} & no & physical & partial & 4 & extended memory & no \\ \hline
			Deng et al. \cite{Deng2010}, \cite{Deng2012} & \begin{tabular}[c]{@{}l@{}}Leon3\\ (softcore)\end{tabular} & no & physical & no & 1-32 & tag TLB & no \\ \hline
			Heo et al. \cite{Heo2015} & \begin{tabular}[c]{@{}l@{}}Leon3\\ (softcore)\end{tabular} & no & physical & no & 1 & \begin{tabular}[c]{@{}l@{}}packed array \\ (bitmap)\end{tabular} & no \\ \hline
			Lee et al. \cite{Lee2016} & \begin{tabular}[c]{@{}l@{}}Leon3\\ (softcore)\end{tabular} & no & physical & partial & 1 & \begin{tabular}[c]{@{}l@{}}packed array\\ (bitmap)\end{tabular} & no \\ \hline
			Wahab et al. \cite{Wahab2017} & ARM Cortex-A9 & no & N/A & no & 1-32 & address and tag & no \\ \hline
			This work & ARM Cortex-A9 & yes & virtual & yes & 1-32 & TMMU & yes \\ \hline
		\end{tabular}
	}
\end{table}

Table~\ref{table:related_work_comparison} provides a comparison of this work with previous off-core approaches that use a dedicated co-processor to monitor the applications executed on the main CPU. Implementations for off-core approaches are generally done using softcores (e.g.~Leon3 or SPARC V8 architecture) as the main CPU. In such solutions, the design is not easily portable on hardcores. Moreover, some designs do not target modern OS such as Linux. Therefore, it is necessary to overcome these limitations to implement DIFT.\\ 

This work is an improvement of \cite{Wahab2017} which proposes to use ARM CoreSight debug components, static analysis, and instrumentation to recover required information for DIFT on ARM SoCs. However, the solution proposed in \cite{Wahab2017} has several limitations. The coprocessor is implemented using a MicroBlaze softcore which limits its performance because the softcore needs to fetch instructions from memory, decode the instruction in software and then compute tags. Furthermore, it lacks kernel support, tagging of floating point and multi-threaded applications cannot be used.\\

The contributions of this paper are the following:
\begin{itemize}
	\item Flexible security policies implementation in hardware: previous off-core solutions lack ways of specifying security policies (compile-time only or runtime only) and do not offer support for multiple security policies of different tag granularities (page, word, etc.).
	\item This work proposes to tag virtual addresses instead of physical addresses as done in related works. It offers the possibility to use the proposed approach with modern OS and hardcore CPUs with memory management units.
	\item This work is compatible with multi-threaded applications mainly thanks to the context ID information obtained from CoreSight PTM. It allows filtering trace using Context ID and to determine the exact order of operations on the CPU for each thread.
	\item This work explains how the Linux kernel communicates with the FPGA fabric available in a Zynq device (including a dual-core Cortex-A9 and reconfigurable logic): information missing in most, if not all, existing works. It shows how each information container is tagged and how the kernel synchronizes with the DIFT coprocessor. 
	\item This work tracks all information flows unlike most existing works. It allows detecting a wider range of attacks, when compared to \cite{Wahab2017}, including attacks targeting library code. Furthermore, the communication time overhead is improved by a factor 3.8 compared to existing work strategy, described in \cite{Heo2015}, with similar software and hardware constraints.
\end{itemize}

\section{Proposed architecture}\label{proposed-architecture}

\subsection{Overall architecture}\label{dift-coprocessor-global-design}

\begin{figure}[htbp]
	\centering
	\includegraphics[width=.8\textwidth]{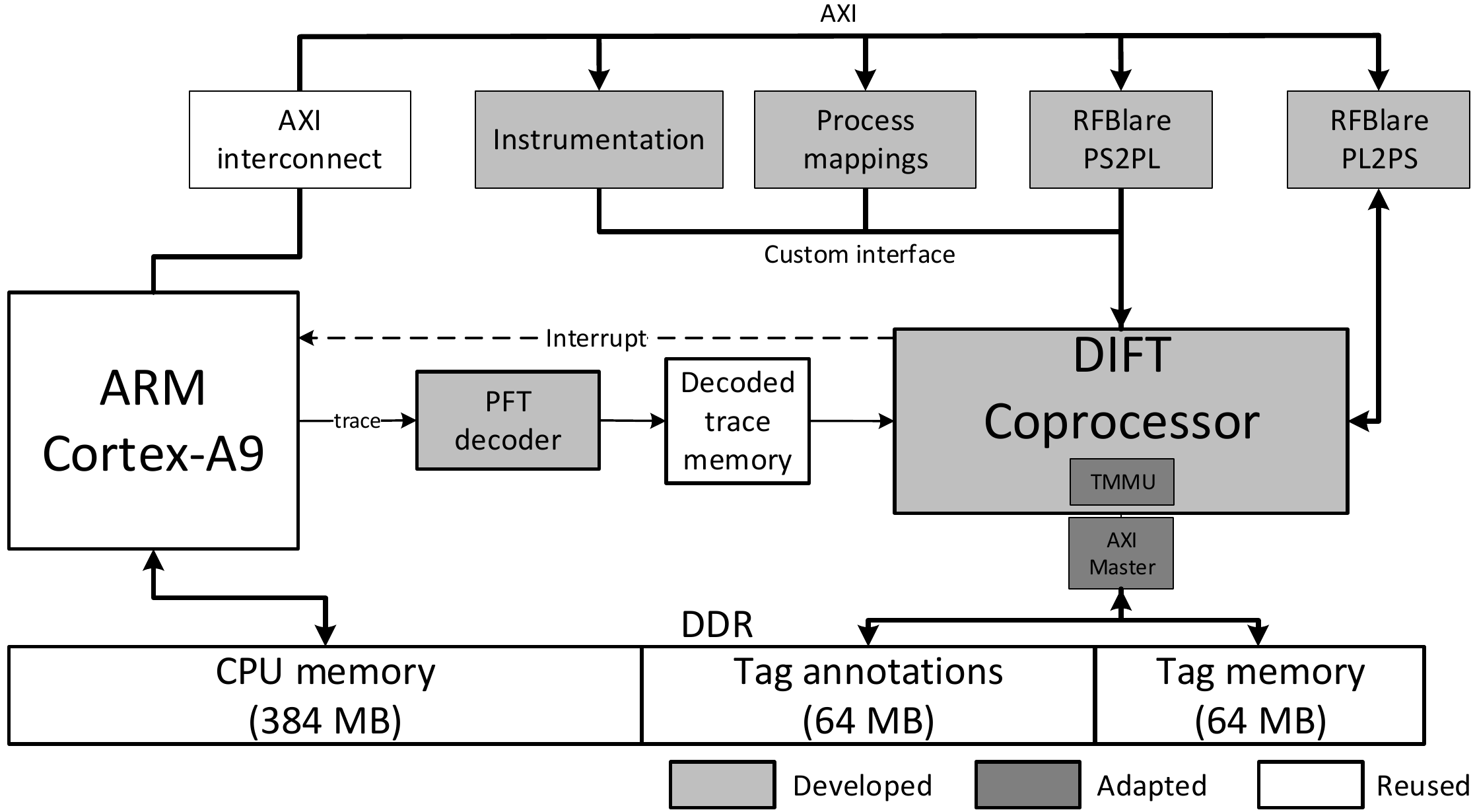}
	\caption{Overall system design with a DIFT coprocessor}
	\label{img:coprocessor_global}
\end{figure}

Figure~\ref{img:coprocessor_global} shows the overall architecture of the proposed approach. The architecture is inspired from an existing off-core approach \cite{Wahab2017}. Instead of using a Microblaze as DIFT coprocessor as in \cite{Wahab2017}, this work takes advantage of a custom DIFT coprocessor. Other main differences are flexible security policies implementation, virtual memory tagging and support for multi-threaded software.\\ 

This paragraph briefly explains how IFT operations are managed by the overall architecture. To begin with, the application is compiled using a custom embedded Linux distribution and SDK (\emph{Software Development Kit}) generated using Yocto toolchain \cite{Yocto}. The modified compiler instruments the application binary to add \texttt{str} instructions as explained in Figure \ref{img:instrumeneted_binary}. During compilation, static analysis is done in order to generate annotations shown in Table \ref{table:annotations}. When the user launches the application, the Linux kernel loads application sections (e.g. \texttt{.text} section) in memory and annotations in tag annotations memory section. The kernel configures the CoreSight PTM to trace the entire \texttt{.text} section of the application. The trace comes out via the EMIO (\emph{Extended Multiplexed Input/output}) interface in PFT (\emph{Program Flow Trace}) protocol as described in \cite{ARMCoreSightPFT}. The PFT decoder decodes the trace and stores it in the decoded trace memory.

The decoded trace only contains information on starting address of basic blocks. What information flows happen inside a basic block needs to be determined. Static analysis allows to recover information flows for each basic block in the application and store them in tag annotations section of memory in a similar way as described in \cite{Wahab2017}. However, some information, such as \texttt{load}/\texttt{store} addresses, cannot be resolved statically. This missing information is obtained through instrumentation via Instrumentation IP (Figure~\ref{img:coprocessor_global}). The DIFT coprocessor requires information from other modules such as process mappings IP which is explained in sections \ref{tag-memory-management} and \ref{process-mappings}. RFBlare IPs, described in section \ref{kernel-support}, are used to communicate with the Linux kernel in order to get the tag of memory address or to set the tag of memory address. The DIFT coprocessor is responsible for managing these IPs, computing and storing tags in tag memory section.

\subsection{Flexible security policies using DIFT coprocessor}\label{flexible-security-policies}
The DIFT coprocessor has been designed in order to provide maximum flexibility in terms of security policy specification. The security policy specifies how to propagate and check tags in order to detect a particular kind of attack \cite{Hritcu2015}. There are two missing aspects in existing works regarding security policies: flexibility to specify security policies and tag granularities.\\

Security policies can be specified at compile-time or at runtime. The compile-time solution~\cite{Heo2015} consists of hard-coding the propagation and checking operations using dedicated opcodes, during the compilation of the application. For instance, consider that the security policy states that all arithmetic and logic instructions on ARM core result in the logical \texttt{OR} operation on their corresponding tags. Then, during static analysis, for all arithmetic and logical instructions of the program code, an \texttt{OR} operation is hard-coded to compute the tags of their operands. The runtime solution~\cite{Dalton2007} requires a special register called TPR (\emph{Tag Propagation Register}) to specify the operation that has to be done on tag values. This time, the static analysis gives the operands for each instruction and the class of ARM instruction: Arithmetic/logical, Load/Store, Branch, Floating point Load/Store. Thanks to class information, the DIFT coprocessor can determine with the help of the TPR register, the operation that needs to be done to propagate the tags corresponding to the instruction executed on the main CPU. The main advantage is that the TPR value can be modified at runtime to modify the policy without recompiling the application.\\ 

No existing works provide the flexibility to specify security policies using both methods. This work proposes an architecture that can implement either one of these approaches providing developers more flexibility to implement security policies. The DIFT coprocessor ISA (\emph{Instruction-Set Architecture}) has two different types of instructions: specific instructions for compile-time method and specific instructions for runtime method. The last ones are used in combination with TPR and TCR (\emph{Tag Check Register}) \cite{Dalton2007} to implement a runtime security policy. Table \ref{table:annotations} sums up different types of instructions (called annotations) supported by the custom DIFT coprocessor described in this work. There are four annotation types: Tag initialization, Tag ALU, Tag Load/Store and compound annotations. The same set of annotations is also included for the floating-point code. Tag initialization annotations can be used in both runtime and compile-time methods to initialize tags of registers or memory addresses. Tag ALU annotations propagate tags for registers. For instance, \texttt{TagRRR} annotation (on the fourth row of Table \ref{table:annotations}) shows a runtime annotation that contains \texttt{type} field and operands \texttt{T1, T2,} and \texttt{T3}. The operation \texttt{op} for this annotation will be determined, at runtime, by reading TPR register value for the corresponding \texttt{type}. If the \texttt{type} field is arithmetic and the TPR register states that an \texttt{AND} operation must be done on source operands to compute destination tag, then the operation done on TMC (\emph{Tag Management Core}) unit of DIFT coprocessor is \texttt{T1 = T2 AND T3}. 

\begin{table}[htbp]
	\centering
	\caption{Overview of DIFT coprocessor (TMC) instructions (called annotations)}
	\label{table:annotations}
	\resizebox{\textwidth}{!}{%
	\begin{tabular}{|l|l|l|l|p{6.8cm}|}
		\hline
		\textbf{Instruction type} & \textbf{Opcode}       & \textbf{Operation type} & \textbf{Example annotation} & \textbf{Action} \\ \hline
		\multirow{3}{*}{Tag initialization} & \texttt{TagRImm} & irrelevant & \texttt{TagRImm T1,\#1000} & \texttt{T1} = 1000 \\ \cline{2-5}
		& \texttt{TagRR} & irrelevant & \texttt{TagRR T2,T1} & \texttt{T2} = \texttt{T1} \\ \cline{2-5}
		& \texttt{TagMR} & irrelevant & \texttt{TagMR R1,T1} & Mem[\texttt{R1}] = \texttt{T1} \\ \hline
		\multirow{2}{*}{Tag ALU} & \texttt{TagRRR} & runtime & \texttt{TRR type T1,T2,T3} & \texttt{T1} = \texttt{T2} op \texttt{T3} \\ \cline{2-5}
		& \texttt{TagRRR2} & compile-time & \texttt{TagRRR2 AND T1,T2,T3} & \texttt{T1} = \texttt{T2} AND \texttt{T3} \\ \hline
		\multirow{4}{*}{Tag Load/Store} & \texttt{TagMTR} & runtime & \texttt{TagMTR type R1,T1,\#4} & Mem
		[\texttt{R1}+4] = \texttt{T1} \\ \cline{2-5}
		& \texttt{TagTRM} & runtime & \texttt{TagTRM type T1,R1,\#4} & \texttt{\texttt{T1} = Mem[\texttt{R1}+4]} \\ \cline{2-5}
		& \texttt{TagMTR2} & compile-time & \texttt{TagMTR2 R1,T1,\#4} & Mem[\texttt{R1}+4] = \texttt{T1} \\ \cline{2-5}
		& \texttt{TagTRM2} & compile-time & \texttt{TagTRM2 T1,R1,\#4} & \texttt{T1} = Mem[\texttt{R1}+4] \\ \hline
		\multirow{6}{*}{Compound} & \texttt{TagITR} & runtime & \texttt{TagITR T3,T1,\#4} & Mem[TMMU(Instrumentation)] = \texttt{T3}\\ \cline{2-5}
		& \texttt{TagTRI} & runtime & \texttt{TagTRI T4,T2,\#4} & \texttt{T4} = Mem[TMMU(Instrumentation)] \\ \cline{2-5}
		& \texttt{TagITR2} & compile-time & \texttt{TagITR2 T12,\#4} & Mem[TMMU(Instrumentation + 4)] = \texttt{T1} \\ \cline{2-5}
		& \texttt{TagTRI2} & compile-time & \texttt{TagTRI2 T2,\#4} & \texttt{T2} = Mem[TMMU(Instrumentation + 4)] \\ \cline{2-5}
		& \texttt{TagKTR} & compile-time & \texttt{TagKTR T1} & Mem[RFblare\_PL2PS] = \texttt{T1} \\ \cline{2-5}
		& \texttt{TagTRK} & compile-time & \texttt{TagTRK T2} & \texttt{T2} = TMMU(Mem[RFblare\_PS2PL]) \\ \hline
	\end{tabular}
}
\end{table}

Furthermore, existing off-core approaches do not provide different tag sizes and use a fixed tag size of one bit in most cases. Therefore, security policies that require multiple bits for a tag (such as heap overflow detection~\cite{Hritcu2015}) cannot be implemented. This work offers hardware support for a tag size up to 32 bits. In addition, it can support multiple security policies as discussed in Section~\ref{case-studies}.

\subsection{Tag virtual memory}\label{tag-virtual-memory}
In related works, tags are associated with physical memory addresses. However, Linux applications are compiled to use virtual addresses and the MMU is responsible to translate them into physical addresses during execution. Existing solutions consider that it is possible to recover translation information from the MMU. This assumption is only realistic  if the main CPU is a softcore tightly coupled with the DIFT co-processor. On a hardcore, a solution could be to modify the Linux kernel in order to send to the coprocessor information about PTEs (\emph{Page Table Entries}), which are managed by the kernel. However, this can be costly because each time a translation is done, the virtual and physical page numbers need to be sent to the FPGA part as well. This information can be difficult to obtain from the kernel and needs lots of minor modifications to the kernel source code which is constantly changing. In addition, the trace generated by the PTM contains virtual addresses. Therefore, to limit the kernel modifications and to avoid virtual to physical address translation overhead, a tag is associated to a virtual address rather than a physical address.

\subsection{Tag memory management}\label{tag-memory-management}

We rely on a TMMU (\emph{Tag Memory Management Unit}) to associate tags to a memory region used by the main CPU. The TMMU, see Figure~\ref{img:coprocessor_global}) is functionally similar to an MMU used in CPUs. It translates each process virtual address to a physical tag address i.e. the physical address which contains the tag associated to the virtual address. It is implemented as an associative array of 64 entries with some additional logic. Each entry contains a virtual page number and the corresponding physical page number where the tag is located. In order to initialize the TMMU, the mappings of different segments of the process (e.g. code and data) are sent to the FPGA (Process mappings IP in Figure~\ref{img:coprocessor_global}). This information is retrieved while the application is being loaded by the kernel and is sent to the FPGA part via process mappings IP (Figure~\ref{img:coprocessor_global}). The process mappings IP contain 64 registers to store information for virtual address page numbers.

\subsection{Coprocessor design}\label{coprocessor-design}

\begin{figure}[htbp]
	\centering
	\includegraphics[width=.8\textwidth]{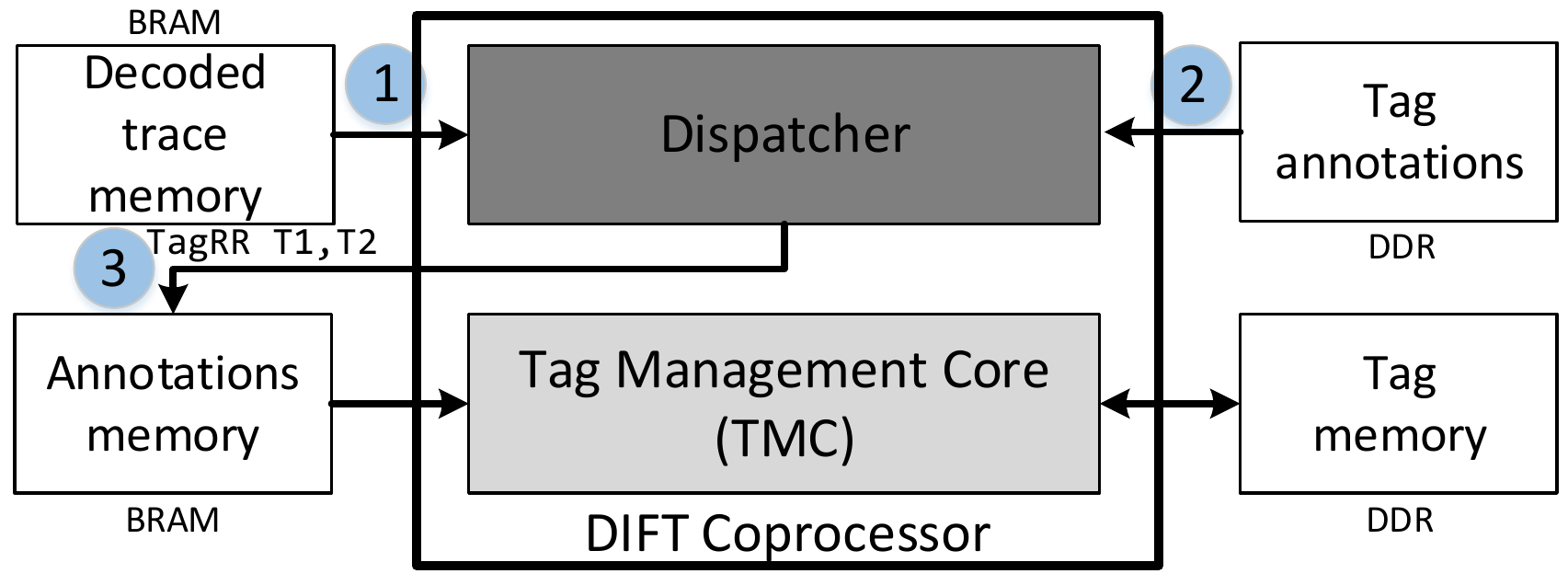}
	\caption{Coprocessor architecture \label{img:dift_coprocessor}}
\end{figure}

Figure~\ref{img:dift_coprocessor} illustrates the high-level architecture of the DIFT coprocessor. The coprocessor has two main submodules: the dispatcher and the TMC unit.

\begin{figure}[htbp]
	\centering
	\includegraphics[width=.8\textwidth]{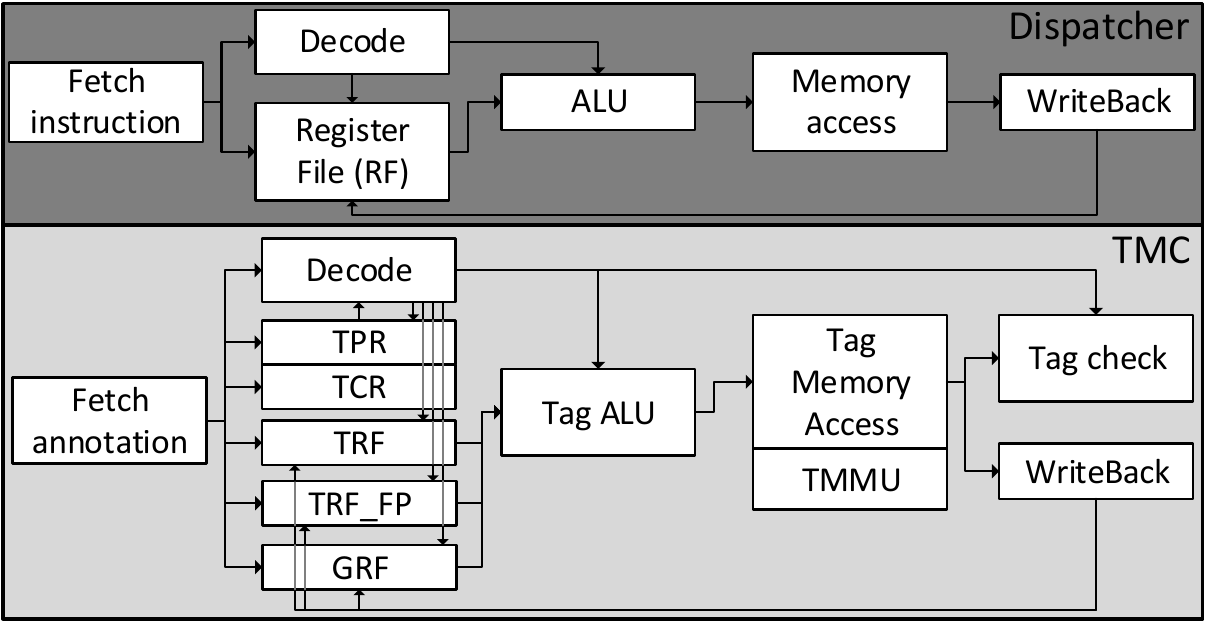}
	\caption{DIFT coprocessor microarchitecture \label{img:dift_coprocessor_microarchitecture}}
\end{figure}

The dispatcher initializes and manages all IPs. Its main objective is to find annotations in program execution order (thanks to decoded trace) and to store them in local annotations memory for the TMC unit. The dispatcher is implemented as a classical five stages pipelined MIPS CPU (Figure~\ref{img:dift_coprocessor_microarchitecture}) that allows executing general operations. It reads decoded trace (1), finds annotations corresponding to the decoded trace by reading tag annotations memory section (2) and stores them in local annotation memory (3).\newpage

The Figure~\ref{img:dift_coprocessor_microarchitecture} shows the internal architecture of the DIFT coprocessor. The TMC core is in charge of propagating and checking tags according to a security policy. The TMC core contains 3 register files: TRF, TRF\_FP, and GRF. TRF contains tags corresponding to the main ARM core registers (\texttt{r0} to \texttt{r15}) while TRF\_FP contains tags for the ARM floating point registers (\texttt{s0} to \texttt{s31}). GRF contains general purpose registers. If a compile-time security policy is being implemented, the tag check operation is performed by Tag ALU. Otherwise, at runtime, the tag check module is responsible for checking tags at the writeback stage. If a tag check fails, then an interrupt is sent to the ARM core to trigger a counter-measure (e.g. stopping the application).\\

The DIFT coprocessor managed two different types of data values: tag values, stored in TRF (\emph{Tag Register File}) and general values, stored in GRF (\emph{General Register File}). TRF is required in order to store tags of ARM CPU registers and GRF is required for general purpose computation. For instance, if a tag of memory address needs to be set to a value, then a general purpose register is needed to store the memory address.\\

The TMC core is responsible for decoding annotations and computing
specified tag operations either in the annotation itself or by using the security
policy registers. It is a dedicated coprocessor for decoding annotations
and is pipelined in five stages (Fetch, Decode, Execute, Memory Access, Write Back). It executes annotations that are detailed in Table \ref{table:annotations}. This core can be duplicated in order to deal
with multiple security policies or in order to propagate tags for
multiple threads or processes.

\subsection{Multi-thread support}\label{multi-thread-support}
To the best of our knowledge, no existing hardcore solutions support multi-threaded applications. The main issue in existing architectures is how the FPGA can determine which thread is running on the main CPU. This work proposes to take advantage of the trace generated by the PTM in order to determine the exact order of the thread execution. The ARM CoreSight PTM can be configured, providing OS support, to recover the TID (\emph{Thread ID}) of threads on each context switch. By filtering the trace with TID, the execution order of threads can be determined. As the trace itself respects the execution order of each thread~\cite{Wahab2017}, the CFG (\emph{Control Flow Graph}) of each thread can be determined. In other words, the DIFT coprocessor can determine, thanks to the context ID feature, the thread executed on the ARM CPU and can propagate tags for the corresponding information flows. As each thread requires a dedicated TMC unit, the number of threads supported by the proposed architecture is limited by the number of TMC units that can be implemented in FPGA.

\subsection{Software requirements}\label{software-requirements}

All existing off-core approaches require software modifications. However, existing works do not provide enough information to easily reproduce their work. Therefore, in this subsection, all proposed software modifications are explained in order to make this work reproducible.

\subsubsection{Instrumentation}

The DIFT coprocessor can reconstruct the program execution path by creating a CFG using the decoded trace. However, it has no information on what happens inside each CFG basic block. The output of the static analysis, as done in \cite{Wahab2017}, states how to propagate tags for each CPU instruction. However, for memory instructions, the static analysis cannot compute the memory addresses that are only known at runtime.

\begin{figure}[htbp]
	\centering
	\subfloat[Original application]{
		\includegraphics[width=.4\linewidth]{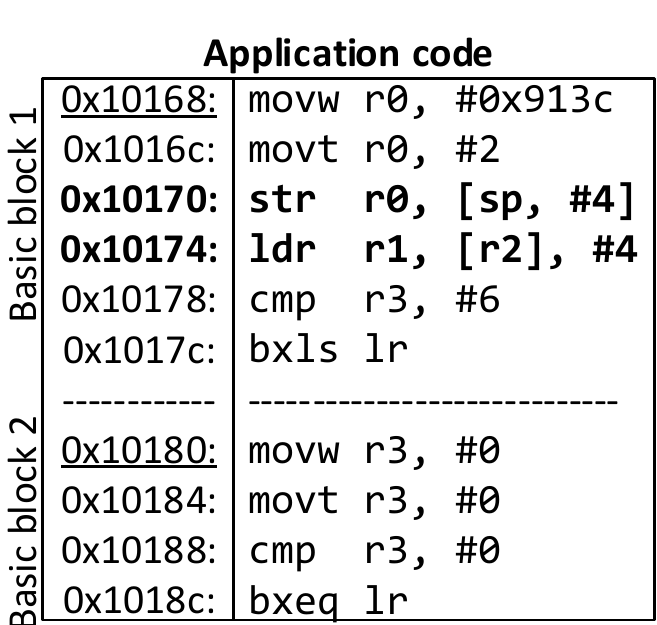}
		\label{img:instrumentation_original}
	}
	\subfloat[Instrumented application]{
		\includegraphics[width=.4\linewidth]{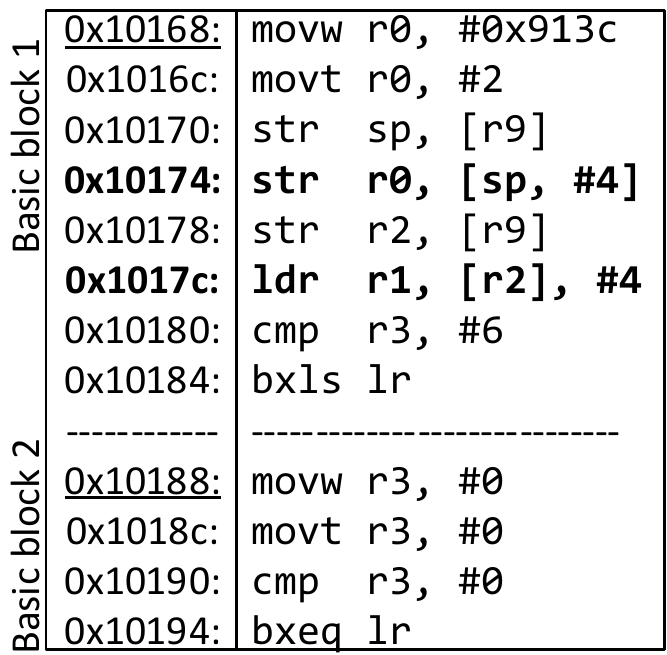}
		\label{img:instrumentation_instrumented}
	}
	\caption{Binary instrumentation using modified LLVM}
	\label{img:instrumeneted_binary}
\end{figure}

This situation is illustrated in Figure~\ref{img:instrumeneted_binary}. Figure~\ref{img:instrumentation_original} shows the original application code. The starting address of basic blocks 1 and 2 (underlined addresses) are recovered from the decoded trace. For memory instructions (in bold in Figure~\ref{img:instrumentation_original}) such as \texttt{ldr} and \texttt{str} at addresses \texttt{0x10170} and \texttt{0x10174} of the original application, the value of registers (\texttt{sp} and \texttt{r2}) cannot be computed from static analysis.\\

This information is needed by the DIFT coprocessor to propagate tags from a register to a memory address (in case of a store operation) or from a memory address to a register (in case of a load operation). In this work, this information is obtained by instrumenting the original application binary. The instrumented application is shown in Figure~\ref{img:instrumentation_instrumented}. Before each memory instruction, another instruction is added (store instructions at addresses \texttt{0x10170} and \texttt{0x10178} of Figure~\ref{img:instrumentation_instrumented}) that sends the missing register value to the memory address contained by \texttt{r9}. The \texttt{r9} register has been reserved and is not used by the application. It contains the virtual address associated with the physical address of instrumentation IP (shown in Figure~\ref{img:coprocessor_global}) such that any store to \texttt{r9} results in a write to the instrumentation IP.

\subsubsection{Kernel support}\label{kernel-support}
Existing hardware DIFT approaches only handle tags associated to RAM and registers. However, they do not take into account information stored on mass storage, i.e. tags associated to files. This feature is important to handle inter-process communications and data persistence (after a reboot). Moreover, users are more inclined to specify a policy at the file level rather than at the memory address or register level. For example, a user can easily identify the files that are supposed to contain confidential information, such as passwords.\\

Handling file implies some kernel support. However, the major kernel modifications are limited to file Input/Output interface. Only a limited number of system calls such as \texttt{read} or \texttt{write} have to be modified. To handle tags associated to files, we rely on RFBlare~\cite{Georget2017}. RFBlare is a modified Linux kernel that implements an OS-level DIFT monitor. This monitor saves tags as file meta-data using file system extended attributes. RFBlare original behavior consists in propagating tags from files to the process memory whenever a file is read and from memory to files whenever a file is written. This OS-level approach is coarse-grained since only one tag is used to abstract the whole memory of a process.\\

We modify RFBlare to disable the OS-level tag propagation and we only used the feature that allows associating tags to files. We add some code to enforce a communication between RFBlare and the DIFT co-processor for each file I/O. For instance, when a \texttt{read} system call occurs on the ARM core, RFBlare allows retrieving the tag of the file being read, the address of the buffer where the read data is stored and the number of bytes read. These three values are sent by the kernel to the FPGA part using RFBlare\_PS2PL FIFO IP (shown in Figure~\ref{img:coprocessor_global}). Similarly, if a program writes to a file (e.g. \texttt{write} system call), the kernel sends the memory address of the buffer being written and the size of the buffer. Then, the DIFT coprocessor fetches the corresponding tag and sends it back to the kernel. RFBlare uses this tag to set the new tag of the file. All the communication for the \texttt{write} system call use RFBlare\_PL2PS FIFO IP (shown in Figure~\ref{img:coprocessor_global}). These FIFOs have an AXI-Lite interface to communicate with the ARM core and a custom FIFO interface to communicate with the TMC.\\

The main CPU and the DIFT coprocessor has to be synchronized since the main core runs faster than the DIFT coprocessor on FPGA. In this work, this synchronization is done thanks to the  RFBlare\_PL2PS and  RFBlare\_PS2PL FIFO mechanisms. As the ARM core and the DIFT coprocessor runs on different frequencies, the attack might be detected after execution of the malicious code. However, it will not compromise the system because an attack needs a system call (for example \texttt{write}) to damage the system. However, each system call waits for the DIFT coprocessor to finish tag computation before carrying on execution. Therefore, the synchronization mechanism makes sure that the software attack does not affect the system. 

\subsubsection{Process mappings}\label{process-mappings}

The DIFT coprocessor needs to know memory mappings of the program in order to properly initialize tag MMU. We modified the ELF binary loader (\texttt{binfmt\_elf.c}) to send memory mapping of the process. This way, the DIFT coprocessor starts by initializing TMMU before any other operation is done.

\section{Case studies}\label{case-studies}

This section shows two use cases where the proposed architecture can outperform existing works: handling multiple security policies and multi-threaded applications.\\

This work takes advantage of the internal architecture of the DIFT coprocessor to reuse developed modules. Figure~\ref{img:dift_coprocessor_multithreaded} shows the proposed architecture for analyzing two threads. The hardware consists of a single dispatcher and two TMC units. This design can also be used to configure multiple security policies. This work proposes to duplicate some modules of the DIFT coprocessor to track multiple threads. The other possible solution is to use context switch i.e. each time the ARM processor switches to another thread, the context of DIFT coprocessor is saved and the new context is restored. However, this solution will add an important storage overhead and time overhead because it would require storing the entire content of three register files and other important CPU registers.

\subsection{Multiple security policies}\label{multiple-security-policies}

In order to implement multiple security policies of different tag sizes, Dhawan et al. proposed to modify the internal architecture of the CPU to handle data and tag~\cite{Dhawan2015}. Their approach relies on a 128-bit CPU with 64-bit of data and 64-bit of tag. However, their architecture is not modular and flexible. For instance, the area and power overhead remain important even if we consider only a single security policy.\\

\begin{figure}[htbp]
	\centering
	\includegraphics[width=.7\linewidth]{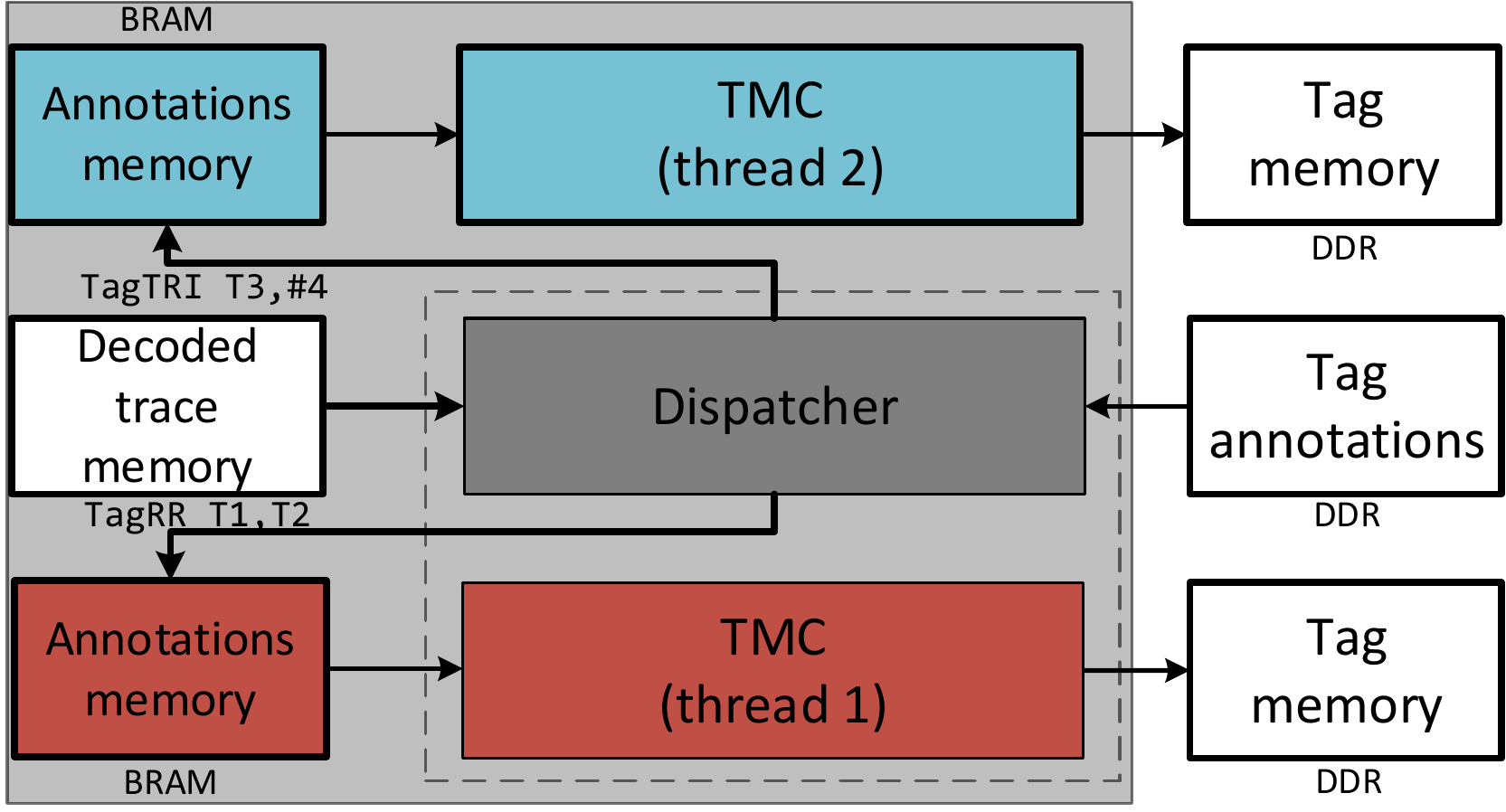}
	\caption{Coprocessor architecture for two threads \label{img:dift_coprocessor_multithreaded}}
\end{figure}

The architecture illustrated by Figure~\ref{img:dift_coprocessor_multithreaded} can be used for implementing multiple security policies. In this case, the thread 1 core executes one security policy and the thread 2 core executes another security policy. The dispatcher reconstructs the execution path of the application and stores annotations in both annotations memories. Then, each TMC unit propagates tags according to its own security policy. If a violation occurs in one of the TMC units, an interrupt is sent to the ARM core to stop execution.\newpage

This approach is modular since we can adapt the number of TMC units to the number of policies to verify in parallel.

\subsection{Multiple threads}\label{multiple-threads-on-a-single-core}

Real-world applications use multiple threads in order to speed up execution. However, existing DIFT mechanisms are not able to track multiple threads because the interface between the main processor and the DIFT coprocessor does not export information that allows determining the thread being executed on the CPU. This approach exploits CoreSight components to 
extract the context ID, which comprises the TID (\emph{Thread ID}) and the ASID (\emph{Application-Specific ID}).

\begin{figure}[htbp]
	\centering
	\includegraphics[width=\linewidth]{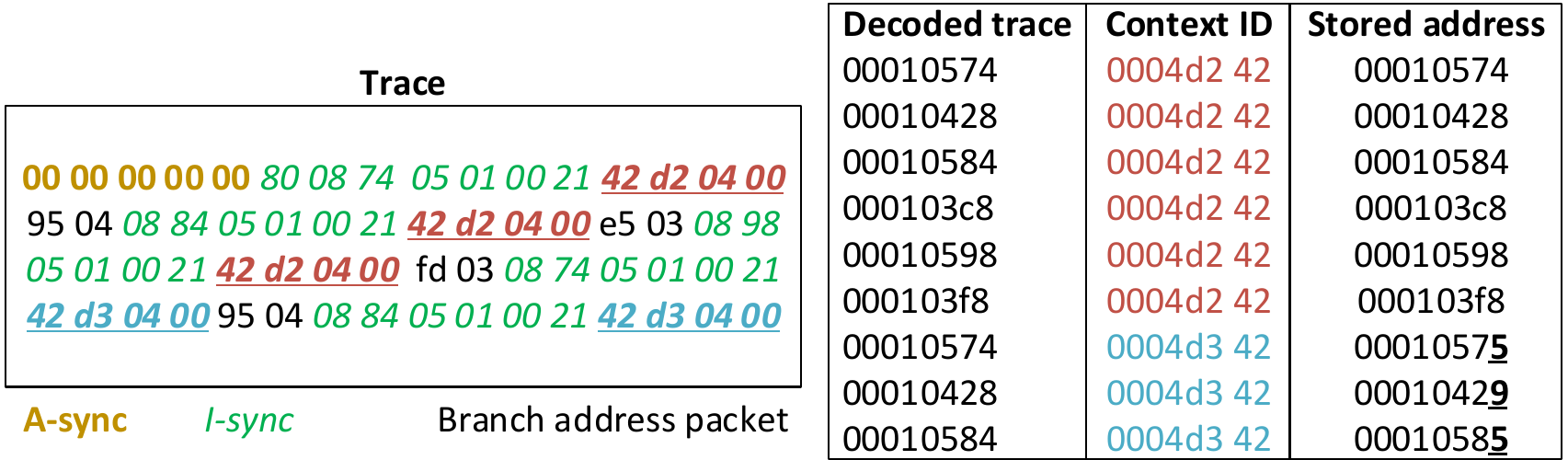}
	\caption{Decoded trace for multiple threads\label{img:multiple_process_trace}}
\end{figure}

\paragraph{Single core}\label{single-core-system}
If two threads are considered and both run on the same CPU core, the context ID field, retrieved by decoding the PTM trace, allows to determine which thread is currently being executed. Figure \ref{img:multiple_process_trace} shows the format of decoded traces in memory. The trace contains I-sync packet (in green) which contains 4 bytes of context-ID (underlined). It can be noticed that the trace contains the same ASID value (\texttt{42}) and two different TID values (\texttt{4d2} and \texttt{4d3}) corresponding to each thread.

The PFT decoder computes the branch address where the ARM core has jumped. These addresses are 4-byte aligned (in ARM state which is the only state allowed to be used during compilation) which means that two bits (0 and 1) are always equal to 0.\\

The PFT decoder also recovers the context ID field and uses these unused two bits to specify whether the decoded trace is generated by the first thread or the second thread. Figure \ref{img:multiple_process_trace} shows that for the second thread, the stored addresses in decoded trace memory are not 4-bit aligned due to the storage of context ID in last two bits. The value of the context ID still needs to be stored in separate registers so that if an attack is detected, the interrupt routine gets the TID of the program in order to kill the process in charge of generating unauthorized behavior. 

\paragraph{Multi-core system}\label{multi-core-system}
If two threads are launched on two different CPUs, the same architecture can be used to propagate tags. However, the trace configuration and the PFT decoder needs to be adapted. In this experiment, we only monitor a single ARM Cortex-A9 core. If the second core is to be considered, the second PTM needs to be configured as well to trace the program. In terms of configuration, when both PTMs are enabled, each PTM needs to insert a trace ID packet so that the funnel \cite{ARMCoreSightTRM} can merge traces onto a single bus.\\

Furthermore, trace sinks (TPIU or ETB) must enable formatting to differentiate trace from different sources. Therefore, the PFT decoder should be adapted in order to consider formatted data instead of raw data considered in this work. The rest of the architecture remains the same as it is similar to multiple threads on a single core case. For multi-core case, a tracing overhead will appear due to trace formatting that adds low overhead of 6\% \cite{ARMCoreSightTechGuide} and an overhead of one byte every time the trace bus switches between trace sources.

\section{Implementation Results}\label{implementation-results}

Xilinx tools 2017.1 are used on a Xilinx Zedboard with a Z-7020 SoC (dual-core Cortex-A9 running at 667 MHz and an Artix-7 FPGA) to implement the architecture shown in Figure \ref{img:coprocessor_global}. The Clang compiler has been used with customized LLVM pass to get the binary and annotations (instructions for TMC). The evaluation described in this section has the following goals: 
\begin{itemize}
	\item Evaluate feasibility of proposed architecture.
	\item Evaluate the cost of software modifications (execution time overhead and memory footprint).
	\item Compare efficiency with related works.
\end{itemize}

\subsection{Area results}\label{area-results}
Table \ref{tab:results} shows the area overhead of this work. Most of the FPGA area is filled by AXI interconnect (5.14\%), dispatcher (4.18 \%) and TMC (3.45\%).\\ 

The overall design takes 20.4\% of the FPGA area. If two security policies are required at the same time, the design would be modified as shown in Figure \ref{img:dift_coprocessor_multithreaded}. The overall design, in case of two security policies, would take additional 4095 slice LUTs, 9074 slice registers (i.e. 8\% additional FPGA logic) and 6 BRAM tiles. In other words, the proposed design can run more than 8 security policies or protect more than 8 processes at the same time.\\

\begin{table}[htbp]
	\begin{center}
		\caption{Post-synthesis area results on Xilinx Zynq Z-7020}
			\begin{tabular}{|l|ccc|}
				\hline
				\textbf{IP Name} & \textbf{Slice LUTs (in \%)} & \textbf{Slice Registers (in \%)}  & \textbf{BRAM Tile} \\ \hline
				Dispatcher & 2223 (4.18\%) & 1867 (1.75\%) & 3 \\ \hline 
				TMC & 1837 (3.45\%) & 2581 (2.43\%) & 6 \\ \hline
				PFT Decoder & 121 (0.23\%) & 231 (0.22\%) & 0 \\  \hline 
				Instrumentation & 676 (1.27\%) & 2108 (1.98\%) & 0 \\ \hline
				Blare PS2PL & 662 (1.24\%) & 2106 (1.98\%) & 0 \\ \hline
				Blare PL2PS & 62 (0.12\%) & 56 (0.05 \%) & 0 \\ \hline
				Decoded trace memory & 0 & 0 & 2 \\ \hline
				AXI Master & 858 (1.61\%) & 2223 (2.09 \%) & 0 \\ \hline
				TMMU & 295 (0.55\%) & 112(0.10 \%) & 3 \\ \hline 
				AXI Interconnect & 2733 (5.14\%) & 2495 (2.34 \%) & 0\\ \hline
				Miscellaneous & 1381 (2.6\%) & 2160 (2.03\%) & 0 \\ \hline 
				\multirow{1}{*}{\textbf{Total Design}} & \textbf{10848 (20.39\%)} & \textbf{15939 (14.98\%)} & \textbf{14 (10\%)} \\ \hline
				\textbf{Total Available} & \textbf{53200} & \textbf{106400} & \textbf{140}\\  
				\hline
			\end{tabular}
		\label{tab:results}
	\end{center}
\end{table}

\subsection{Time overhead analysis}\label{performance-overhead}
CoreSight components do not add any execution time overhead~\cite{Wahab2017}. Even though this work uses a different configuration of CoreSight components since the context ID tracing is enabled, CoreSight components still do not add any noticeable execution time overhead. The static analysis does not add time overhead because it is performed offline, during compilation. Therefore, the timing overhead of the proposed approach is only due to instrumentation. The overhead of instrumentation is analyzed by measuring instrumented application execution time normalized to the original application (non-instrumented) running time. Execution times are measured using a set of custom applications, such as FFT (Fast Fourier Transform) and CRC (Cyclic Redundancy Check) computations, on the Linux kernel 4.9, patched with CoreSight TPIU driver, using the Linux \texttt{perf} command.

\begin{figure}[htbp]
	\centering
	\includegraphics[width=\textwidth]{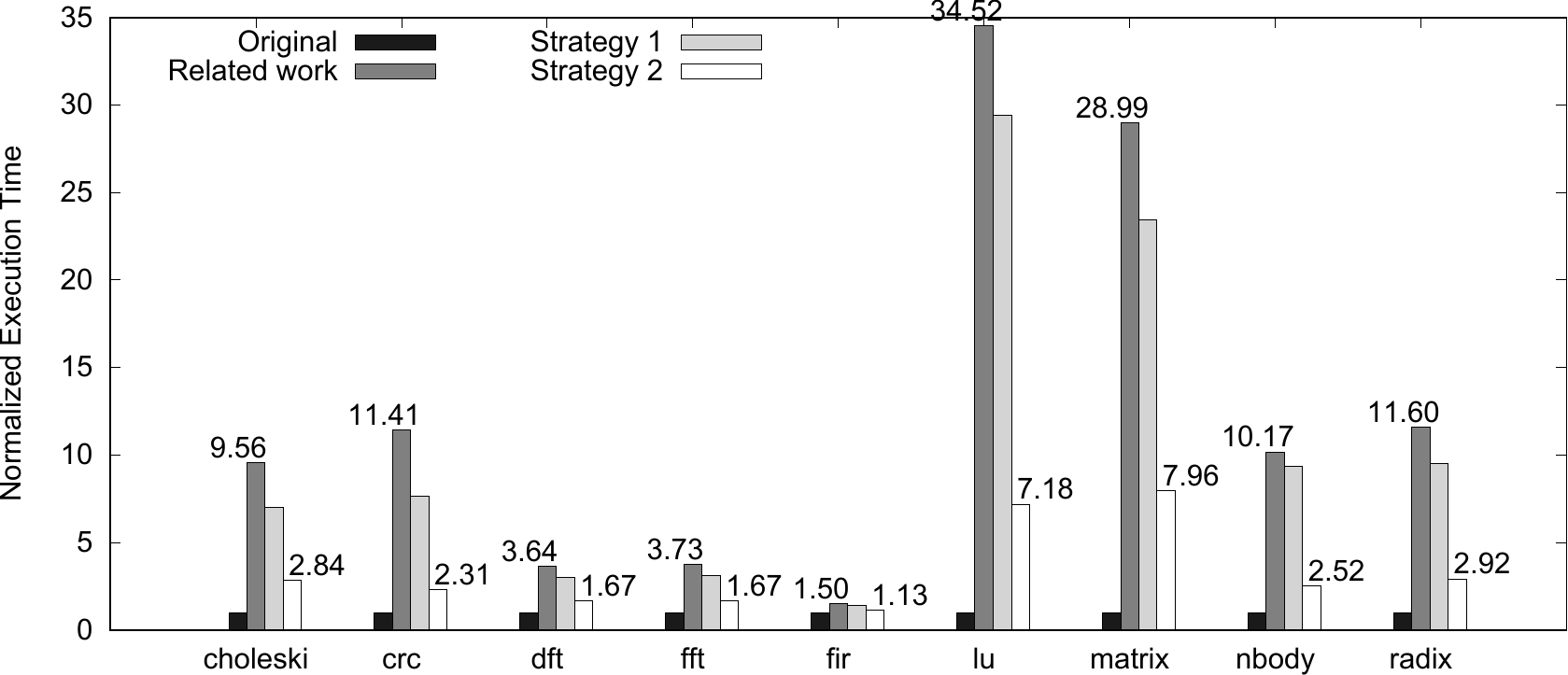}
	\caption{Average execution time overhead for custom benchmark\label{img:instrumentation_time_overhead}}
\end{figure}

Figure \ref{img:instrumentation_time_overhead} shows the average normalized execution time overhead for instrumented application binaries. The related work implementation strategy is detailed in \cite{Heo2015}. Two instrumentation strategies (strategy 1 and 2) are adapted from \cite{Wahab2017}. However, the difference between their proposed strategies and this work resides in static analysis, instrumentation and hardware modules. All information flows are considered in this work unlike \cite{Wahab2017}. Library code is instrumented unlike \cite{Wahab2017} and different configuration of Coresight components is used which changes the design of hardware modules such as PFT decoder.\\ 

Figure \ref{img:instrumentation_time_overhead} shows that if related work strategy as in \cite{Heo2015} is used, the instrumentation overhead, on average, is 12.79 times higher than the original execution time. The instrumentation strategy 1, in which all memory instructions are instrumented, adds significant execution time overhead (on average 10.43 times higher than the original execution time). The instrumentation overhead is high for applications that require more memory operations (such as \texttt{lu} and \texttt{matrix}). However, if only register-relative (other than \texttt{PC, SP,} and \texttt{FP}) memory instructions are instrumented, as in strategy 2, the average communication time overhead is reduced by a factor 3.8 to achieve 3.35 times higher execution time on average. The three main reasons why this overhead remains high are: this work targets hardcore CPU which is not the case in most existing works such as in \cite{Deng2010,Heo2015,Dalton2007,Dhawan2015}, the static analysis considers all information flows rather than function level information flows as in \cite{Heo2015,Wahab2017} and it also instruments library code used by the applications unlike \cite{Wahab2017}.

\subsection{Memory Footprint}

\begin{figure}[htbp]
	\centering
	\includegraphics[width=.8\textwidth]{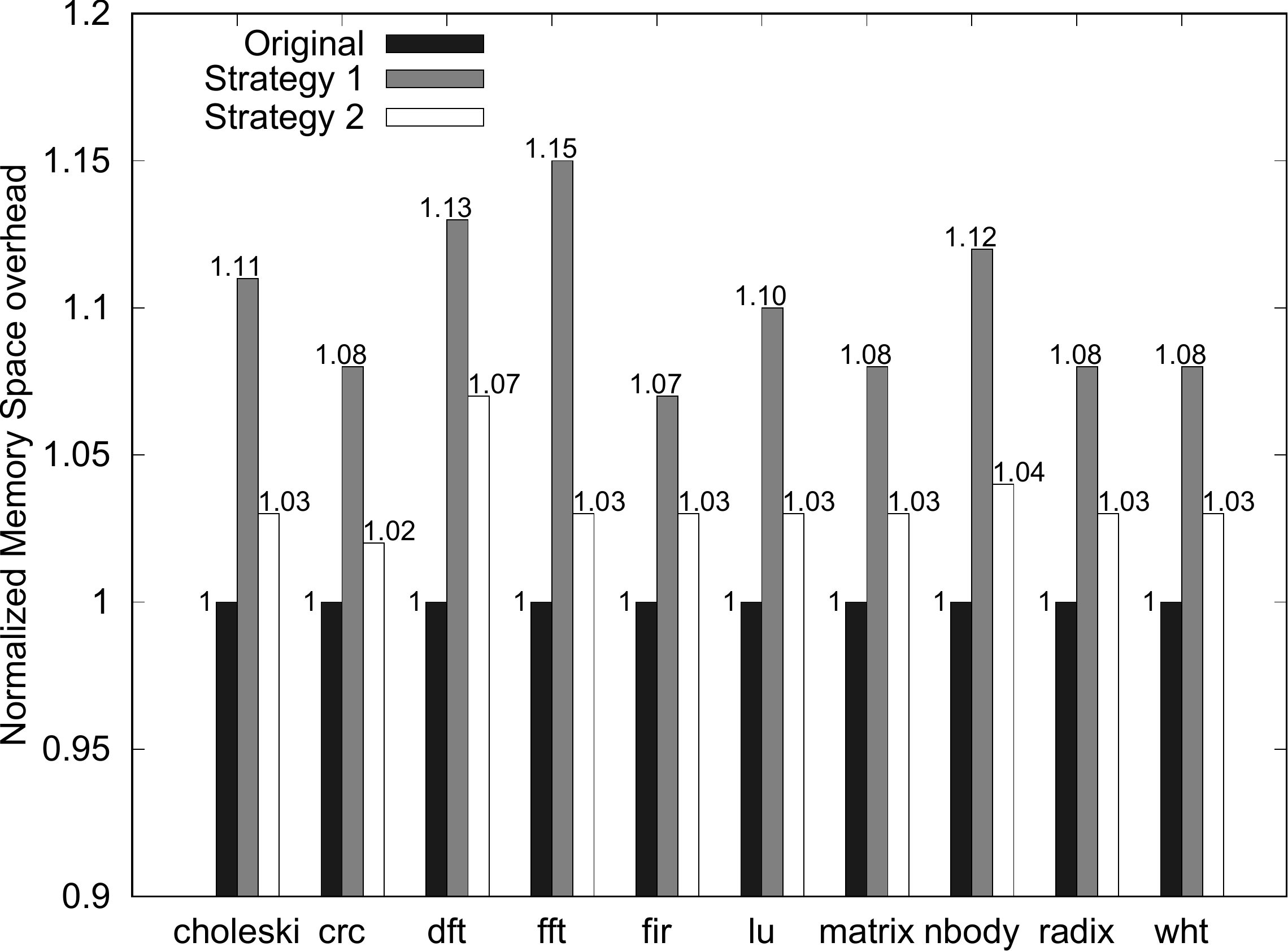}
	\caption{Memory footprint of custom benchmark\label{img:memory_footprint}}
\end{figure}

Figure \ref{img:memory_footprint} shows the memory space overhead of custom benchmark applications using both strategies.

Binaries are statically compiled i.e. all the code executed by the application is inside the code section of the binary. Therefore, results obtained here take into account all modifications of user code and the library code. The instrumentation strategy 1 adds in average 10\% of memory space overhead and instrumentation strategy 2 adds in average only 3\% of memory space overhead.

\subsection{Comparison with related works\label{subsec:comparison-with--related-works}}
Table \ref{tab_related} shows the comparison of an off-core approach like this work with previous off-core approaches. Comparison of this work with in-core solutions is not done because they are not hardcore portable and are very invasive contrary to the off-core solution used in this work.

\begin{table}[htbp]
	\centering
	\caption{Performance comparison with previous off-core approaches}
		\begin{tabular}{|c|cccccc|}
			\hline
			\textbf{Approaches} & \textbf{Kannan \cite{Kannan2009}} & \textbf{Deng \cite{Deng2010}} & \textbf{Heo \cite{Heo2015}} & \begin{tabular}{@{}c@{}}\textbf{Heo \cite{Heo2015}}\\\textbf{adapted}\end{tabular} & \textbf{Wahab \cite{Wahab2017}} & \textbf{This work}\\\hlineB{5}
			Area overhead     & 6.4\%    & 14.8\%   & 14.47\%  & N/A  & \textbf{0.47\%}   & 0.95 \% \\ \hline
			Power overhead    & N/A      & \textbf{6.3\%}    & 24\%     & N/A     & 16\%     & 16.2\% \\ \hline
			Max frequency     & N/A      & \textbf{256 MHz}  & N/A      & N/A      & 250 MHz  & 250 MHz\\ \hline
			\begin{tabular}{@{}c@{}}Communication\\time overhead\end{tabular} & N/A      & N/A      & 60\%         & 1280\%       & \textbf{5.4\%}    & 335\%\\\hlineB{5}
			\begin{tabular}{@{}c@{}}Hardcore\\portability \end{tabular}  & No       & No       & \textbf{Yes} & \textbf{Yes} & \textbf{Yes}      & \textbf{Yes} \\\hline
			Main CPU          & Softcore & Softcore & Softcore & \textbf{Hardcore} & \textbf{Hardcore} & \textbf{Hardcore}\\\hline
			\begin{tabular}{@{}c@{}}Library\\instrumentation\end{tabular} & N/A      & N/A      & partial  & \textbf{Yes}      & No       & \textbf{Yes} \\\hline
			\begin{tabular}{@{}c@{}}All\\information flows\end{tabular}   & No & No & No & \textbf{Yes} & No & \textbf{Yes} \\\hline
			FP support        & No       & No       & No       & No       & No       & \textbf{Yes}\\\hline
			\begin{tabular}{@{}c@{}}Multi-threaded \\ support \end{tabular} & No       & No       & No       & No       & No       & \textbf{Yes}\\ \hline
		\end{tabular}
	\label{tab_related}
\end{table}

Compared to \cite{Wahab2017}, this work has higher area overhead because it provides support for FP and multi-threaded software that requires additional modules in the FPGA area. Furthermore, the power overhead of this work is similar to \cite{Wahab2017} because this work does not use any DSPs unlike the coprocessor used in \cite{Wahab2017} which is a MicroBlaze with DSP units that are more power consuming than normal FPGA logic. The maximum frequency achievable is comparable to existing works. The most important difference resides in communication time overhead. It may appear that this overhead is higher than values reported in related works.\newpage

The fifth column, Heo \cite{Heo2015} adapted, shows that if the proposed instrumentation strategy in related work \cite{Heo2015} is adapted with the same static analysis and information flows constraints as in this work, then the communication time overhead obtained on Zynq architecture reaches 1280\%. However, the proposed solution in this work can reduce this overhead to 335\% by a factor $\frac{1280}{335} = $ 3.8. The best communication time overhead reported in existing works is 5.4 \% (\cite{Wahab2017}) but their proposed solution lacks support for all information flows and library instrumentation. Furthermore, the reported value is an estimation based on the number of instrumented instructions whereas the value reported in this work is obtained on the Zedboard platform and the time is measured using \texttt{perf} tool on Linux kernel v4.9. It is very important to consider all information flows and instrument libraries because it allows detecting an important range of attacks unlike most existing works. For instance, a simple attack on existing works could be to add a wrapper around library function and use Linux kernel dynamic \texttt{LD\_PRELOAD} feature to avoid detection of any malicious library code. However, this work is able to detect the execution of malicious library code as its tracking is not ignored as in most existing works. This work provides support for floating point (through additional instructions and additional register file) and multi-threaded software thanks to context ID feature of the CoreSight PTM as described in section \ref{case-studies}: support for these features is missing in related works.\newpage

\section{Conclusion}\label{conclusion}
This work is the first work providing support and flexibility to implement multiple security policies which can either be specified at run-time or compile-time. It provides features missing in related works: protection of floating point code and multi-threaded software. Both features are essential in protecting real-world applications. This work takes advantage of CoreSight components along with LLVM modifications to implement information flow security policies. The communication time overhead is reduced to more than 380\% if compared to existing work strategy \cite{Heo2015}. Area results show interesting perspectives in terms of implementing multiple security policies and protecting multiple processes. For instance, protecting two threads requires another TMC unit which adds an additional area overhead of 8\%. As this work deals with multi-threaded applications that do not share memory, the next step will be to include support for threads that share memory. Another improvement would be adding support for more than 8 threads by making context switches on each TMC unit.

\bibliographystyle{unsrt}
\bibliography{IEEEabrv}

\begin{thebibliography}{10}

\bibitem{Dalton2007}
Michael Dalton, Hari Kannan, and Christos Kozyrakis.
\newblock Raksha: A flexible information flow architecture for software
  security.
\newblock volume~35, pages 482--493, New York, NY, USA, June 2007. ACM.

\bibitem{Kannan2009}
H.~Kannan, M.~Dalton, and C.~Kozyrakis.
\newblock Decoupling dynamic information flow tracking with a dedicated
  coprocessor.
\newblock In {\em 2009 IEEE/IFIP International Conference on Dependable Systems
  Networks}, pages 105--114, June 2009.

\bibitem{Heo2015}
Ingoo Heo, Minsu Kim, Yongje Lee, Changho Choi, Jinyong Lee, Brent~Byunghoon
  Kang, and Yunheung Paek.
\newblock Implementing an application-specific instruction-set processor for
  system-level dynamic program analysis engines.
\newblock {\em ACM Trans. Des. Autom. Electron. Syst.}, 20(4):53:1--53:32,
  September 2015.

\bibitem{Dhawan2015}
Udit Dhawan, Catalin Hritcu, Raphael Rubin, Nikos Vasilakis, Silviu Chiricescu,
  Jonathan~M. Smith, Thomas~F. Knight, Jr., Benjamin~C. Pierce, and Andre
  DeHon.
\newblock Architectural support for software-defined metadata processing.
\newblock {\em SIGARCH Comput. Archit. News}, 43(1):487--502, March 2015.

\bibitem{Wahab2017}
M.~A. Wahab, P.~Cotret, M.~N. Allah, G.~Hiet, V.~Lapôtre, and G.~Gogniat.
\newblock Armhex: A hardware extension for dift on arm-based socs.
\newblock In {\em 2017 27th International Conference on Field Programmable
  Logic and Applications (FPL)}, pages 1--7, September 2017.

\bibitem{ARMCoreSightETM}
{ARM}.
\newblock {\em {Embedded Trace Macrocell Architecture Specification}}.

\bibitem{Lee2016}
Jinyong Lee, Ingoo Heo, Yongje Lee, and Yunheung Paek.
\newblock Efficient security monitoring with the core debug interface in an
  embedded processor.
\newblock {\em ACM Trans. Des. Autom. Electron. Syst.}, 22(1):8:1--8:29, May
  2016.

\bibitem{Fytraki2014}
S.~Fytraki, E.~Vlachos, O.~Kocberber, B.~Falsafi, and B.~Grot.
\newblock Fade: A programmable filtering accelerator for instruction-grain
  monitoring.
\newblock In {\em 2014 IEEE 20th International Symposium on High Performance
  Computer Architecture (HPCA)}, pages 108--119, February 2014.

\bibitem{Chen2008}
Shimin Chen, Michael Kozuch, Theodoros Strigkos, Babak Falsafi, Phillip~B.
  Gibbons, Todd~C. Mowry, Vijaya Ramachandran, Olatunji Ruwase, Michael Ryan,
  and Evangelos Vlachos.
\newblock Flexible hardware acceleration for instruction-grain program
  monitoring.
\newblock In {\em Proceedings of the 35th Annual International Symposium on
  Computer Architecture}, ISCA '08, pages 377--388, Washington, DC, USA, 2008.
  IEEE Computer Society.

\bibitem{Nagarajan2008}
Vijay Nagarajan, Ho-Seop Kim, Youfeng Wu, and Rajiv Gupta.
\newblock Dynamic information flow tracking on multicores.
\newblock {\em Workshop on Interaction between Compilers and Computer
  Architectures, (colocated with HPCA)}, February 2008.

\bibitem{Deng2010}
Daniel~Y. Deng, Daniel Lo, Greg Malysa, Skyler Schneider, and G.~Edward Suh.
\newblock Flexible and efficient instruction-grained run-time monitoring using
  on-chip reconfigurable fabric.
\newblock In {\em Proceedings of the 2010 43rd Annual IEEE/ACM International
  Symposium on Microarchitecture}, MICRO '43, pages 137--148, Washington, DC,
  USA, 2010. IEEE Computer Society.

\bibitem{Deng2012}
D.~Y. Deng and G.~E. Suh.
\newblock High-performance parallel accelerator for flexible and efficient
  run-time monitoring.
\newblock In {\em IEEE/IFIP International Conference on Dependable Systems and
  Networks (DSN 2012)}, pages 1--12, June 2012.

\bibitem{Yocto}
{Yocto project}.
\newblock {\em {Yocto Project Reference Manual}}.

\bibitem{ARMCoreSightPFT}
{ARM}.
\newblock {\em {CoreSight Program Flow Trace Architecture Specification}}.

\bibitem{Hritcu2015}
A.~A. d.~Amorim, M.~Dénès, N.~Giannarakis, C.~Hritcu, B.~C. Pierce,
  A.~Spector-Zabusky, and A.~Tolmach.
\newblock Micro-policies: Formally verified, tag-based security monitors.
\newblock In {\em 2015 IEEE Symposium on Security and Privacy}, pages 813--830,
  May 2015.

\bibitem{Georget2017}
Laurent Georget, Mathieu Jaume, Guillaume Piolle, Fr{\'e}d{\'e}ric Tronel, and
  Val{\'e}rie Viet~Triem Tong.
\newblock Information flow tracking for linux handling concurrent system calls
  and shared memory.
\newblock In Alessandro Cimatti and Marjan Sirjani, editors, {\em Software
  Engineering and Formal Methods}, pages 1--16, Cham, 2017. Springer
  International Publishing.

\bibitem{ARMCoreSightTRM}
{ARM}.
\newblock {\em {CoreSight Components Technincal Reference Manual}}.

\bibitem{ARMCoreSightTechGuide}
ARM.
\newblock {\em CoreSight Technology System Design Guide}.

\end{thebibliography}
\end{document}